\documentclass[a4paper,oneside,10pt]{article}
\usepackage[T1]{fontenc}
\usepackage{a4wide}
\usepackage{amssymb}
\usepackage[arrow,matrix]{xy}

\newcommand{\C}{\mathbb{C}}
\newcommand{\R}{\mathbb{R}}
\newcommand{\cH}{\mathcal{H}}
\newcommand{\cS}{\mathcal{S}}
\newcommand{\tens}{\otimes}
\newcommand{\xD}{\mathcal{D}}

\title{The general boundary approach\\ to quantum gravity}
\author{Robert Oeckl\footnote{Centre de Physique Th\'eorique,
CNRS Luminy,
13288 Marseille, France, email: oeckl@cpt.univ-mrs.fr}\\ \\
Talk to be given at ICP2004, Tehran, Iran, January 2004}
\date{CPT-2003/C.4619\\ November 25, 2003}

\begin{document}
\maketitle

\begin{abstract}
We present an approach to quantum gravity based on the general
boundary formulation of quantum mechanics, path integral quantization,
spin foam models and renormalization.
\end{abstract}

\section{Introduction}

Even after decades of intensive research the problem of reconciling
quantum mechanics and genereal relativity remains an elusive one. One
of the core difficulties faced by many approaches is the \emph{problem of
  time} (see Isham's excellent review \cite{Ish:problemoftime}). This
may be viewed as arising from the contradiction of a classical time
variable underlying standard quantum mechanics and the supposed
quantum nature of time emerging out of a quantum theory of general
relativity. Another problematic issue is the fact that natural
observables in classical general relativity are highly non-local -- a
consequence of the role of diffeomorphisms as gauge
symmetries. Translated into a quantum theory of 3-geometries this
would presumably imply that natural observables are at least spatially
non-local. Consequently, describing local experiments in the quantum
theory would be a priori difficult.

A fundamental reason for these difficulties is arguably the way
standard quantum mechanics is formulated. One presumes a classical and
fixed space-time to set up the formalism.
Even special relativity only
enters ``through the back door'' in quantum field theory and is not
manifest in the formalism.
The assumption of standard approaches to quantum general relativity is
then that the quantization of space and time might be considered in
a second step, resting on the formalism thus set up.

The ``general boundary'' approach is precisely aimed at addressing
the abovementioned difficulties. On the one hand it avoids the problem
of time and on the other hand it is inherently compatible with general
covariance and allows for local descriptions of local experiments. The
price to pay is that we need to modify or rather extend the standard
formalism of quantum mechanics. The key idea is that transition
amplitudes are associated with regions of space-time and states are
associated with their boundaries. This reduces in the case of regions
that are time intervals extended over all of space essentially to the
standard formalism.

This approach was introduced in \cite{Oe:catandclock,Oe:boundary} and
is currently under development. We present some of its basic ideas and
motivations in Section~\ref{sec:boundary}. In
Section~\ref{sec:pathint} we show how the path integral affords a
natural guide to quantization in this context. In
Section~\ref{sec:road} we go a step further and propose a blueprint
for defining models of quantum gravity of such a type. The latter
incorporates ideas from loop quantum gravity \cite{Rov:lqg}, spin foam
models \cite{Per:sfmodels} and renormalization \cite{Oe:renormdisc}.

\section{General boundaries}
\label{sec:boundary}

The general boundary formalism is an extension of the standard
formalism of quantum mechanics with the aim of natural compatibility
with general covariance. It should be
applicable in particular to quantum general relativity. The extension
as compared to the standard formalism might be sketched as follows. In
the standard formalism one associates a Hilbert space of states with
each time-slice of a global foliation of
space-time. An evolution takes place between two
such time-slices and is represented by a unitary operator. Associated
with states in the two time-slices is a transition amplitude, whose
modulus square determines the probability of finding the final state
given that the initial one was prepared.

More explicitly, we have Hilbert spaces
$\cH_1$ and $\cH_2$ of states associated to the initial time $t_1$ and 
final time $t_2$. The evolution is described by an operator
$U(t_1,t_2):\cH_1\to \cH_2$. The transition amplitude for an initial state
$\psi_1\in\cH_1$ to evolve into a final state $\psi_2\in\cH_2^*$ is
written as $\langle \psi_2 | U(t_1,t_2) | \psi_1\rangle$.

In terms of a space-time
picture an evolution operator is associated to a region of space-time,
namely the product of the time interval with all of space. The states
(initial and final) are naturally associated with the boundary
components of this region. Indeed, the first step of our our
generalization consists of
``forgetting'' the a priori distinction between initial and final
state. Instead we consider a state space which is the tensor product
of the two state spaces associated with the time-slices. The new state
space is naturally associated with the boundary as a whole.
That such a formulation is consistent is a rather non-trivial fact. It
crucially relies on a symmetry of quantum field theory coming out of
the LSZ reduction that allows to exchange individual
particles between the initial and final state without changing the
amplitude. (Of course, a CPT transformation must be performed on the
particle at the same time and phase space measures do change.)

Formalizing this, we have a generalized state space $\cH_{[t_1,t_2]}$
that is the
tensor product $\cH_1\tens \cH_2^*$. A state $\psi$ in
$\cH_{[t_1,t_2]}$ is a
(linear combination of) tensor product(s) $\psi_1\tens \psi_2$ of
states in $\cH_1$ and $\cH_2$. The transition amplitude is then a map
$\cH_{[t_1,t_2]}\to\C$ which we denote by $\rho_{[t_1,t_2]}$. In terms
of the conventional 
notation this means
\begin{equation}
 \rho_{[t_1,t_2]}(\psi_1\tens\psi_2)=\langle \psi_2| U(t_1,t_2)
 |\psi_1\rangle .
\label{eq:ampl}
\end{equation}

In terms of a measurement process, the initial and final state are both
encoded in a state of the generalized state space. The evolution operator
becomes a linear map from the generalized state space to the
complex numbers, associating transition amplitudes to generalized
states. The second step is to generalize from the special regions of
space-time that are time intervals extended over all of space to more
general regions. For this to be consistent we need to introduce a
composition property. This property requires that when we glue two
regions of space-time together the evolution map associated with the
composite must equal the composition of the evolution maps associated
with the original pieces. This generalizes the composition of time
evolutions in standard quantum mechanics. More precisely, one
demands the properties of a topological quantum field
theory \cite{Ati:tqft}.

Given a space-time region (4-manifold) $M$ with boundary $\Sigma$, we
write $\cH_\Sigma$ for the state space associated with the
boundary. For the evolution map (or amplitude) we write
$\rho_M:\cH_\Sigma\to \C$. The situation of conventional quantum mechanics
is recovered if $M$ is the product of all of space $\R^3$ with a time
interval $[t_1,t_2]$. $\Sigma$ is then the union of two components
$\Sigma_1\cup \Sigma_2$, each being all of space $\R^3$ times a point
in time. By the axioms of topological quantum field theory this
implies that $\cH_\Sigma$ decomposes into a tensor product of vector
spaces associated to the components $\cH_\Sigma=\cH_{\Sigma_1}\tens
\cH_{\Sigma_2}$. In this way we recover (\ref{eq:ampl}). Note that
in general (especially if $\Sigma$ is connected) there is no natural
decomposition of $\cH_\Sigma$ into a tensor product and thus no longer
any natural distinction between preparation and observation in quantum
mechanics. This has profound interpretational implications, e.g.\ with
regards to the ``collapse of the wave function''
\cite{Oe:catandclock}.

It might seem that we still suppose an a priori fixed space-time in
contrast to the desire of seeing it emerge from the quantization. However,
this is not really the case. What we do presume is only the topology,
but not the geometry. The geometry and thus the dynamical degrees of
freedom of gravity are really to be encoded in the state. Furthermore, we
might even introduce sums over topologies in the interior and only fix
the topology of the boundary of the space-time region. However, it is
unclear whether this is reasonable.

For a sensible interpretation of measurement processes
we require the boundary of the space-time regions to be
connected. (Note that this is in stark contrast to standard quantum
mechanics.) The
simplest type of region would be one that has the topology of a
4-ball. Indeed, this type of topology might be sufficient for most
practical purposes.

This ``general boundary'' formulation has several advantages:
\begin{itemize}
\item The problem of time disappears. The description of a local
  measurement involves a local region of space-time. Time durations
  in the measurement process make sense if the gravitational degrees
  of freedom of the generalized state are nearly classical. In this
  case time durations (with possible uncertainties) might be read of
  from the state by integrating the metric along paths in the
  boundary, in particular in its time-like parts. The connectedness of
  the boundary is crucial here. 
\item Local measurement processes can be described using a local
  region of space-time only. Neither is recourse made to distant
  events in the universe nor is any knowledge of its global
  structure necessary.
\item It is not necessary to endorse ``realist''
  interpretations of quantum mechanics for this formulation to make
  sense. This is in contrast to other
  approaches which presume for example the existence of a ``wave
  function of the universe''.
\end{itemize}

Since the proposed approach is still in its infancy many issues have
not yet been addressed. Among them is the challenge to supply a
completely satisfactory probability interpretation that does not take
recourse to the special situation of time intervals. Another issue is
that of identifying particle states on general boundaries. For this
reason it is desirable to understand the approach first in the
context of standard quantum field theory and even non-relativistic
quantum mechanics.
Although its motivation comes from quantum gravity
there is no inherent limitation of the general boundary formulation to
this context. To the contrary, it is rather easy to adapt
it to the situation where a fixed background metric is given. The
suitable modification of the axioms of topological quantum field
theory to this case is straightforward.

An interesting consequence of the approach is already exhibited in
non-relativistic quantum mechanics \cite{Oe:boundary}. Consistency
requires that state
spaces contain states for any number of particles, as in quantum field
theory. The reason may be seen to be a similar one in the two cases: A
Lorentz boost in quantum field theory
can be considered as ``tilting'' the boundary components of a time
interval times all of space so that they do no longer (from the point
of view of the original frame) correspond to constant times. Along the
same lines,
indistinguishability of particles in the general boundary approach
leads to pair creation and annihilation.

\section{Path integral quantization}
\label{sec:pathint}

So far we have only described the general formalism that we would like
quantum theories to be formulated in. In this section we will consider
the issue of quantization. That is, we describe heuristically how to
obtain quantum theories of the general boundary type from classical
theories. It turns out that the path integral approach is naturally
suited for this purpose. Our presentation here is necessarily brief
and we refer the interested reader to \cite{Oe:boundary} for details.

Consider a classical field theory with fields $\phi$ and an action
$\cS[\phi]$. (For simplicity, we do not write the indices for different
fields or field components.) A quantization leading to a general
boundary quantum theory may be carried out roughly as follows. For the
boundary $\Sigma$ of some space-time region $M$ we consider the space
$K_\Sigma$ of field configurations on $\Sigma$. We then define
$\cH_\Sigma$ to be the space $C(K_\Sigma)$ of complex valued functions
on $K_\Sigma$. This definition has the required property that if $\Sigma$
decomposes into connected components then $\cH_\Sigma$ decomposes into
a tensor product of spaces associated with the components.
Explicitly, if $\Sigma=\Sigma_1\cup\Sigma_2$, then
$K_\Sigma=K_{\Sigma_1}\times K_{\Sigma_2}$ and consequently $C(K_\Sigma)=
C(K_{\Sigma_1})\tens C(K_{\Sigma_2})$.
The last ingredient is the amplitude $\rho_M:\cH_\Sigma\to\C$ which
can be given as follows,
\begin{equation}
 \rho_M(\psi)=\int_{K_\Sigma} \xD\phi_0\, \psi(\phi_0)
 \int_{\phi|_\Sigma=\phi_0} \xD\phi\, e^{\frac{i}{\hbar} \cS[\phi]} .
\label{eq:quant}
\end{equation}
The outer integral is over field configurations $\phi_0$ on the
boundary $\Sigma$ and the inner integral is over all field
configurations in the interior that reduce on the boundary to the
given configuration $\phi_0$.

One can now check that these definitions satisfy all the required
properties (the topological quantum field theory type axioms). Also
one checks that one recovers the usual quantum field theoretic
transition amplitudes in the conventional case of $M$ being given by a
time interval, i.e.\ $M=\R^3\times
[t_1,t_2]$.

\section{Model of a non-perturbative approach}
\label{sec:road}

In this section we present a possible blueprint for formulating a
quantum theory of gravity in the general boundary
formulation.

Firstly, given suitable variables and an associated action we can
formally write down the quantization in the general boundary
formulation, given by (\ref{eq:quant}). The most conventional form of
this would be
to chose the metric field as variable and the Einstein-Hilbert
action. However, as is well known, writing down a path integral is
more a statement than the solution of a problem. This is
particularly true for general relativity, where the standard
perturbation approach of quantum field theory fails due to
non-renormalizability.

Path integrals of general relativity have been extensively
studied in the context of the \emph{Euclidean Quantum Gravity} approach
championed by Hawking \cite{Haw:pathintqg}. The basic idea is there
that one can
make sense of the path integrals by evaluating them in a euclideanized
setting, i.e.\ with Riemannian metrics. One then relates the result to
the physical setting with Lorentzian signature.

The basic quantities of interest in Euclidean Quantum Gravity
are indeed very similar to those of interest in the general boundary
approach. However, and this is remarkable, their interpretation is
rather different. In Euclidean Quantum Gravity one adheres to standard
quantum mechanics and wishes to construct transition amplitudes
associated with states living on space-like slices of the
universe. To make this problem manageable one adds time-like boundary
conditions ``at spatial infinity'' if the universe is not closed. This
is essentially equivalent to a compactification of the interior
space-time region. Thus, the resulting picture is
mathematically close to the situation of a compact space-time region
with connected boundary as of interest in the general boundary
approach. This leads to the curious situation that what is known in
Euclidean Quantum Gravity as the ``wave function of the universe''
\cite{HaHa:waveuniv} is
formally similar to the amplitude for a generic 4-ball region of
space-time in the general boundary approach. Reinterpreting the
results of Euclidean Quantum Gravity in the light of the present
approach might offer interesting insights. We shall pursue here a
different route, however, but before embarking on it let us take a
closer look at the boundary.

\subsection{Boundary and causal structure}

We consider the generic case of a smooth 4-ball $B$, the boundary
being a 3-sphere $S$. Suppose we have a solution of the Einstein field
equations in the interior that extends to the ``outside'' (we imagine
the 4-ball embedded into some larger space). Necessarily, the boundary
$S$ has both space-like and time-like parts. In the simplest case
there will be two space-like parts (near the poles) and one time-like
(near the equator). Separating these parts there are null regions which
are generically 2-spheres. In more complicated cases there are several
time-like parts and also more space-like parts. This corresponds to
more ``crumpled'' boundaries. We call the separation of the boundary
into regions of the three types space-like, time-like and null a
\emph{causal structure}.

In order to simplify the physical interpretation in the quantum
situation this suggests to restrict the boundary configurations to
conform to given causal structures. This really means incorporating
the causal structure as a background. We would thus no longer have a
``pure'' topological quantum field theory, but have the extra datum of
causal structure given on the boundaries.

Another interesting aspect of this is that we could consider 4-balls
with boundaries that are almost everywhere null except for
submanifolds of dimension at most two (double cone shape). This would
require giving up
smoothness. Indeed, recent work of Reisenberger \cite{Rei:talkcbpf}
shows
that general relativity with null boundaries takes a particularly
simple form and might thus be particularly suitable for
quantization.
Note also that giving up smoothness allows for causal structures where
the boundary is almost everywhere space-like (lens shape) or time-like
(pointed cigar shape).

\subsection{Quantization of the boundary -- ``kinematic'' state space}

We now turn to the ``quantization'' of the boundary, i.e.\ the
construction of the ``kinematic'' state space $\cH_\Sigma$ associated
with the boundary $\Sigma$.
Using the metric as variable the configuration
space of interest should be that of metrics on the boundary. However,
physically significant is really only the \emph{intrinsic} metric on the
boundary. This is because the other components of the metric in a
solution of the field equations can be
arbitrarily modified by diffeomorphisms that leave the boundary
invariant. Thus we take the configuration space $K_\Sigma$ to be the
space of intrinsic metrics on $\Sigma$.

The signature of the intrinsic metric on $\Sigma$ is determined by the
causal structure of $\Sigma$. More precisely, the metric has
signature $(+,+,+)$ on space-like parts, $(+,+,-)$ on time-like parts
and $(+,+,0)$ on null parts. Considering a causal structure $C$
to be a background structure on $\Sigma$ we denote the boundary with
background by $\Sigma_C$. Correspondingly, we have a configurations space
$K_{\Sigma_C}$ associated with $\Sigma_C$ that contains only the intrinsic
metrics on $\Sigma$ respecting $C$. The ``background-free'' configuration
space $K_\Sigma$ is then simply the union over the configuration spaces
with the different causal structures,
\[
 K_\Sigma = \bigcup_C K_{\Sigma_C} .
\]
Taking $\cH_{\Sigma_C}$ to be a suitable space $C(K_{\Sigma_C})$ of
functions on $K_{\Sigma_C}$ we obtain,
\[
 \cH_\Sigma = \bigoplus_C \cH_{\Sigma_C} .
\]

Spaces of metrics are rather complicated and
a path integral over metrics in the interior is rather
difficult to make sense of. We shall continue with a different
choice of variables to describe general relativity. This is the
formulation in terms of a 4-bein (frame field) variable $E$ and an
associated connection $A$ of the Lorentz group. Doing so allows us to
establish a relation to \emph{loop quantum gravity}
\cite{Rov:lqg}. This is a canonical approach to quantum gravity, i.e.\
one starts by splitting space-time into time-slices, associates
Hilbert spaces with the time slices and then constructs an
evolution operator (which is actually a projector) between these
Hilbert spaces.

Relevant for us here is the fact that the problem of constructing the
Hilbert space of loop quantum gravity is very similar to the problem
of finding a suitable state space $\cH$ in the general boundary
approach. Using a suitable Hamiltonian formulation of general
relativity (the Ashtekar formulation) one finds roughly that the
3-bein $E'$ and 3-connection $A'$ induced in the space-like
hypersurface $\Sigma$ are conjugate variables.\footnote{The situation
  is really somewhat more complicated as $A'$ is the Ashtekar
  connection which also includes an extrinsic curvature part. Also
  $E'$ is really
densitized etc. We disregard these details here as they are not
relevant for the present sketchy account.} For the quantization it turns
out to be advantageous to think of the connection $A'$ as ``position''
and of the 3-bein $E'$ as ``momentum''. The configuration space
$K_\Sigma$ associated with $\Sigma$ is thus the space of connections
$A'$ on $\Sigma$. Loop quantum gravity starts by constructing the
Hilbert space $\cH_\Sigma$ as a suitable space $C(K_\Sigma)$ of
functions on $K_\Sigma$.

This is roughly done as follows. Natural gauge invariant functions on
the space of connections are Wilson loops, i.e.\ the holonomy along a
closed loop in $\Sigma$ evaluated with the character of some
irreducible representation of the gauge group. In turns out that
slightly more general functions are more suitable: spin
networks. These are graphs in $\Sigma$ whose edges are labeled by
irreducible representations and whose vertices are labeled by
intertwiners.

The similarity of the problem suggests to construct the state space
$\cH_\Sigma$ associated with the boundary $\Sigma$ of a 4-manifold $M$
in a similar way. The main difference to loop quantum gravity is that
$\Sigma$ is no longer everywhere space-like. This implies that the
gauge group of the connection is no longer everywhere $SO(3)$ (or its
double cover $SU(2)$). Rather, it alternates between $SO(3)$ (for
space-like parts) and $SO(2,1)$ for time-like parts, according to the
signature of the associated metric. For null parts (of codimension
zero) it should even be $SO(2)$.
Thus, the above discussion of the decomposition of the configuration
space $K_\Sigma$ in terms of different causal structures applies
again. The difference is that instead of the signature of the metric
now the gauge group changes in the different parts of the causal
structure.

For $\Sigma$ with a given causal structure $C$ we would have
$\cH_{\Sigma_C}$ spanned by spin networks embedded into $\Sigma$ where
edges carry representations of the gauge group that is associated with
the respective causal part of $\Sigma$ into which they are
embedded. One would then have to establish suitable matching
conditions between the parts of different type.

A simpler approach would be not to restrict the gauge group
from the outset but take spin networks of the Lorentz group $SO(3,1)$
(or its double cover $SL(2,\C)$) everywhere for $\cH_\Sigma$. This
would mean to not restrict to given causal structures. This is also a
priori easier to connect to the construction of the path integral in
the interior that we shall discuss in the following.

Note that all our discussion here is rather sketchy, imprecise and
preliminary. The details would require considerable work.

\subsection{Quantizing the interior -- amplitudes}

As already mentioned, the path integral in the interior of the
4-manifold $M$ that we want to use to define the amplitude
(\ref{eq:quant}) is hard to make sense of. A way to make sense of this
path integral is through \emph{spin foam models}
\cite{Per:sfmodels}. In the present context this might be described as
follows.

As above, we use 4-bein $E$ and connection $A$ variables to describe
the degrees of freedom of general relativity. To make sense of the
path integral we start by introducing a \emph{discretization} of the
space-time $M$. That is, we decompose $M$ into little pieces (cells)
that are 4-balls. Then we draw edges connecting the centers of any
two adjacent 4-balls. As in lattice gauge theory we encode the
connection $A$ by assigning group elements to the edges. What enters
the classical action is the curvature of $F$ which (again as in
lattice gauge theory) we model trough holonomies around faces that are
bounded by the edges of the discretization.
It turns out
that proceeding further along these lines 
allows to write down a
discretized version of the path integral known as a spin foam
model. Roughly speaking, the integral over 4-bein and 
connection fields is transformed to a sum over
assignments of irreducible representations
to the faces of the discretization and assignments of
intertwiners to the edges.

The trouble is that there are many ambiguities in the process of
defining such a spin foam model. Perhaps more seriously, all spin foam
models of this kind have the property that the value of the path
integral thus defined depends on the discretization. We shall come
back to these difficulties in a moment.

Usually spin foam models are considered for manifolds without
boundary. What is interesting to us here is of course the situation
with boundaries. More particularly, we also want a definition of the
outer integral in (\ref{eq:quant}). Here it turns out that the spin
network description on the boundary and the spin foam description in
the interior fit together very well. Indeed one can think of a spin
network as obtained by a cut through a spin foam. Moreover the
structures (representations and intertwiners) associated with the
elements of spin network and spin foam also fit together. Indeed, at
least if we take connections on the boundary to be associated with the
full gauge group (the Lorentz group) the structures fit naturally
together and can be used to provide a definition of the full
expression for the amplitude (\ref{eq:quant}).
If we restrict to subgroups according to a causal structure on the
boundary the situation would become more complicated.
Again, we remain rather sketchy here and spelling out the crucial
details would require considerable work.

\subsection{Renormalization}

Using a spin foam model to define the amplitude (\ref{eq:quant}) we
have the serious problem that the amplitude depends on the unphysical
choice of a discretization. This must be eliminated to have a well
defined model.

There are two main approaches at achieving this: One is to ``sum over
spin foams'' and the other is renormalization. The underlying idea in
the first case is that the discretized theory describes physics
directly at a fundamental length scale. Consequently, a summation over
quantum geometries of different ``sizes'' has to be performed to
capture the path integral.
The problem of this approach is that a ``sum over spin foams'' is not
at all well-defined. Although a proposal for solving this problem
exists \cite{ReRo:stfeyn}
it is rather ad hoc and suffers from the appearance of non-topological
contributions and potential divergences.

We shall pursue the second approach here, renormalization. The
underlying idea here is that the discretization is a kind of cut-off
that has to be removed in the full theory. This is in analogy to the
situation in lattice gauge theory and in line with our above reasoning
for introducing the discretization in the first place.
However, unlike for lattice gauge
theory there is no immediate notion of continuum limit. As there
is no metric there is no quantification of scale that could be used to
define this. However, this is not really necessary.

Even in the case of lattice gauge theory the actual evaluation
(usually by numerical means) of the theory never takes place in the
continuum but always at finite lattice size. The crucial aspect of
renormalization that allows for physically well defined predictions is
that it relates the theory at different scales. Thus, to compute some
physical quantity (like the expectation value of a Wilson loop) it
suffices to evaluate the theory at a scale that is sufficiently small
with respect to the scale of this particular quantity. When we want to
compute a different physical quantity requiring a different scale in
the theory the renormalization group flow tells us how to change the
coupling constants of the theory so that the result will be physically
compatible with the previous one.

Indeed, we can pursue the same idea in the present
setting. Renormalization would need to relate the model defined with
different discretizations. However, now there are no length scales as
we have no background metric. It was proposed in \cite{Oe:renormdisc}
how to extend renormalization methods to this type of
situation. Of course, the model needs to have coupling constants that
can be adjusted for compensating a change of discretization. A crucial
difference to situations in conventional lattice models is however,
that a change of discretization is something much more general than a
global change of scale. In particular, a discretization can change
locally. It was argued in \cite{Oe:renormdisc} that this necessitates
\emph{local} coupling constants, i.e. coupling constants associated to
elements of the discretization. Furthermore, a concrete model for
quantum gravity with such local coupling constants was proposed in
\cite{Oe:renormdisc}, interpolating between the Barrett-Crane model
and BF-theory. We will leave the discussion independent of a
particular model, however.

Let us spell out details of the renormalization: We have
discretizations $D$ of the manifold $M$ and a spin foam model that
assigns values to the path integral (\ref{eq:quant}). Note that we
couple to a spin network on the boundary $\Sigma$ and that it is
convenient to think of this as living in a discretization $\Delta$ of
the boundary that matches the discretization $D$ in the interior.

A discretization $\Delta$ of the boundary $\Sigma$ gives rise to a
restricted state space $\cH_{\Sigma,\Delta}$ that only carries the spin
networks which ``fit'' into $\Delta$.
(Here we leave open the question whether we also
fix a causal structure on the boundary from the outset.)
We have a natural embedding from
$\cH_{\Sigma,\Delta}$ into $\cH_{\Sigma}$ and also between different
discretizations of the boundary $I_{\Delta,\Delta'}:\cH_{\Sigma,\Delta}\to
\cH_{\Sigma,\Delta'}$ if $\Delta'$ is a refinement of $\Delta$. These
maps satisfy the commutative diagram
\[
\xymatrix{\cH_{\Sigma,\Delta}\ar[d] \ar[r] & \cH_\Sigma\\
 \cH_{\Sigma,\Delta'}\ar[ur]} .
\]
Indeed, in the absence of a direct definition of $\cH_\Sigma$ this could
be used as a definition through a limit.

A model of the type under consideration provides an amplitude map
$\rho_{M,D,\lambda}:\cH_{\Sigma,\Delta}\to \C$ for a given
discretization $D$ of $M$ that restricts to a discretization $\Delta$
of the boundary $\Sigma$ and for given values $\lambda$ of the local
coupling constants.
What we are looking for is a map
$\rho_M:\cH_\Sigma\to\C$ that gives a definition of (\ref{eq:quant}).
Thus, we need to get rid of the discretization dependence and of the
dependence on the coupling constants. The latter should be fixed by
desired physical properties. This fixing is done at a given
discretization. The renormalization problem is to determine the
dependence of the coupling constants $\lambda$ on the discretization
$D$, so that the physical properties remain the same. This parallels
the situation in ordinary quantum field theory. There, coupling
constants are determined at a given scale and renormalization is
the determination of the scale dependence, so that physical
properties remain the same.

In the case at hand the physical properties are encoded in the
amplitude map $\rho$. 
An important ingredient here is that for any (spin network) state $\psi\in
\cH_\Sigma$ there exists a discretization $\Delta$ of $\Sigma$ such
that there is a state $\psi_\Delta\in\cH_\Delta$ which is mapped to
$\psi$ under the inclusion $\cH_{\Sigma,\Delta}\to\cH_\Sigma$.
We can thus formulate the renormalization problem as follows.
Determine for any discretization $D'$ that reduces to
$\Delta'$ on
$\Sigma$ the coupling constants $\lambda'$ so that with respect to a
given coarser discretization $D$ (and $\Delta$) and given coupling constants
the following equality holds for any $\psi_\Delta\in
\cH_{\Sigma,\Delta}$:
\begin{equation}
 \rho_{M,D,\lambda}(\psi_\Delta)
 =\rho_{M,D',\lambda'}(I_{\Delta,\Delta'}(\psi_\Delta)) .
\label{eq:renorm}
\end{equation}

This requires some explanations and qualifications. Firstly, equation
(\ref{eq:renorm}) should hold for essentially any pair of
discretization $D$ and refinement $D'$. Only in this way is the full
state space
$\cH_\Sigma$ exhausted. Secondly, we not only looking for solutions of
the equation with fixed coupling constants, but rather for solutions
for any possible choice of couplings. That is, $\lambda'$ in
(\ref{eq:renorm}) is to arise from $\lambda$ through an action
associated with the change of discretization $(D,D')$. More precisely,
this is an action of the groupoid of changes of discretizations. It
gives rise to the \emph{renormalization groupoid}, see
\cite{Oe:renormdisc}. This parallels the situation in lattice gauge
theory where one seeks an action of the group of scale transformations
on the space of coupling constants. Thirdly, what we wrote as an
equation might be required only to be an approximate equality. The
size of deviations would have to be controlled by the relation between
the two discretizations $D$, $D'$. However, this is for the moment a
rather speculative point which we do not pursue.

Another issue is the fact that changing a discretization changes the
number of coupling constants as they are local. In particular, the
number of coupling constants might increase. In this case it would not
be possible give a well defined value to the new coupling constants
on the requirement alone the the physics (encoded in $\rho$) of the
previous discretization be reproduced. In other words, new physical
degrees of freedom appear whose behaviour is not yet fixed.
Thus, the action of the renormalization groupoid often goes only in
one direction, namely that of coarsening of the discretization. This
establishes the direction of the renormalization groupoid flow. Again,
this is somewhat similar to situations in ordinary statistical
physics. 

At first it might seem that solving the system of equations
(\ref{eq:renorm}) is a rather hopeless endeavour, as the space of
discretizations is vast. However, as shown (conjecturally) in
\cite{Oe:renormdisc} there is a suitable set of generators of the
groupoid of changes of discretizations. These are called the
\emph{cellular moves}. As any change of discretization (cellular
decomposition) can be achieved through a sequence of cellular moves,
it is sufficient to establish the system (\ref{eq:renorm}) for pairs
of discretizations that differ by a cellular move. This vastly
simplifies the problem.

There is another aspect to be discussed. How is the discretization
required related to a physical property we want to describe? In the
absence of a proper continuum limit
we need to know what discretization we have to choose to answer a
certain physical question. More precisely, the question is how
fine the discretization has at least to be. In the present context
this is rather clear for states on the boundary. For a given state $\psi$
we need to choose a discretization $\Delta$ which is sufficiently fine
so that there exists a state $\psi_\Delta$ which is mapped to $\psi$
under the inclusion $\cH_{\Sigma,\Delta}\to\cH_{\Sigma}$. For the
interior of $M$ the answer is rather less clear and certainly depends
on the model. In any case, one would expect the ``fineness'' of $D$ to be
directly related to the required ``fineness'' of $\Delta$. In the
other extreme we could envisage models that do not involve any
discretization of the interior, but only one of the boundary.

Given a solution of the renormalization problem we could then fix
coupling constants at a given discretization through physical
properties to obtain the required amplitude map
$\rho_M:\cH_\Sigma\to\C$ and thus complete the theory.

\section{Conclusions}

In this paper we have presented the general boundary approach
\cite{Oe:catandclock, Oe:boundary} to
quantum gravity together with possible steps to a complete theory of
quantum gravity. Key advantages of the general boundary formalism are
compatibility from the outset with general covariance, locality, and
the avoidance of the problem of time. However, many steps remain to be
taken to make this formalism complete and a key role should be played
here by the application to ordinary quantum field theory. This
concerns in particular the probability interpretation and the
identification of particle states. For steps in this direction see
\cite{CDORT:vacuum}.

A heuristic guideline for the quantization of a field theory to obtain
a quantum field theory in the general boundary formulation was given
using the path integral. For more details, see
\cite{Oe:boundary}. Complementary to the path integral an operator
picture is also developed \cite{CoRo:genschroed}. This should help in
particular to connect to canonical quantization approaches such as
loop quantum gravity.

As a more concrete proposal to implement the path integral, spin foam
models were proposed as a definition of the path integral and methods
of loop quantum gravity for making sense of the boundary state
spaces. To deal with the ensuing problem of discretization dependence
we suggested a renormalization procedure, adapting the proposal in
\cite{Oe:renormdisc} to the case at hand. The next step will be to
bring this proposal to life, especially with ``realistic'' models such
as the interpolating one of \cite{Oe:renormdisc}.

\subsection*{Acknowledgements}
This work was supported through a Marie Curie Fellowship of the
European Union.

\bibliography{stdrefs}

\begin{thebibliography}{10}

\bibitem{Ish:problemoftime}
C.~J. Isham, {\em Canonical quantum gravity and the problem of time},
  Integrable systems, quantum groups, and quantum field theories (Salamanca,
  1992), NATO Adv. Sci. Inst. Ser. C Math. Phys. Sci., no. 409, Kluwer,
  Dordrecht, 1993, pp.~157--287.

\bibitem{Oe:catandclock}
R.~Oeckl, {\em Schr\"odinger's cat and the clock: Lessons for quantum gravity},
  Class. Quantum Grav. {\bf 20} (2003), 5371--5380, gr-qc/0306007.

\bibitem{Oe:boundary}
R.~Oeckl, {\em A "general boundary" formulation for quantum mechanics and
  quantum gravity}, Phys. Lett. {\bf B 575} (2003), 318--324, hep-th/0306025.

\bibitem{Rov:lqg}
C.~Rovelli, {\em Loop Quantum Gravity}, Living Rev. Rel. {\bf 1} (1998), 1.

\bibitem{Per:sfmodels}
A.~Perez, {\em Spin foam models for quantum gravity}, Class. Quantum Grav. {\bf
  20} (2003), R 43.

\bibitem{Oe:renormdisc}
R.~Oeckl, {\em Renormalization of discrete models without background}, Nucl.
  Phys. {\bf B 657} (2003), 107--138.

\bibitem{Ati:tqft}
M.~Atiyah, {\em Topological quantum field theories}, Inst. Hautes \'Etudes Sci.
  Publ. Math. (1989), no.~68, 175--186.

\bibitem{Haw:pathintqg}
S.~W. Hawking, {\em The path-integral approach to quantum gravity}, General
  Relativity: An Einstein Centenary (S.~W. Hawking and W.~Israel, eds.),
  Cambridge University Press, Cambridge, 1979.

\bibitem{HaHa:waveuniv}
J.~B. Hartle and S.~W. Hawking, {\em Wave function of the Universe}, Phys. Rev.
  {\bf D 28} (1983), 2960--2975.

\bibitem{Rei:talkcbpf}
M.~P. Reisenberger, Talk at Tenth Marcel Grossmann Meeting on General
  Relativity, CBPF, Rio de Janeiro, July 20--26, 2003.

\bibitem{ReRo:stfeyn}
M.~Reisenberger and C.~Rovelli, {\em Spacetime as a Feynman diagram: the
  connection formulation}, Class. Quant. Gravity {\bf 18} (2001), 121--140.

\bibitem{CDORT:vacuum}
F.~Conrady, L.~Doplicher, R.~Oeckl, C.~Rovelli, and M.~Testa, {\em Minkowski
  vacuum in background independent quantum gravity}, to appear in Phys. Rev. D,
  Preprint gr-qc/0307118.

\bibitem{CoRo:genschroed}
F.~Conrady and C.~Rovelli, {\em Generalized Schroedinger equation in Euclidean
  field theory}, Preprint hep-th/0310246.

\end{thebibliography}
\bibliographystyle{amsordx}

\end{document}